\documentclass[aps,prl,reprint,twocolumn,notitlepage,superscriptaddress,longbibliography]{revtex4}
\usepackage[T1]{fontenc}
\usepackage[utf8]{inputenc}
\usepackage{lmodern}
\usepackage{amsmath,amssymb,amsfonts,bm}
\usepackage{graphicx}
\usepackage{hyperref}
\hypersetup{colorlinks=true,linkcolor=blue,citecolor=blue,urlcolor=blue}
\usepackage{xcolor}
\usepackage[normalem]{ulem}

\newcommand{\dd}{\mathrm{d}}
\newcommand{\ii}{\mathrm{i}}
\newcommand{\ee}{\mathrm{e}}

\newcommand{\UU}{\mathbf{U}}

\begin{document}

\title{Unified Hydrodynamic Analogue of Aharonov-Bohm and Lense-Thirring Effects}

\author{A. Singh}
\affiliation{Nonlinear and Non-equilibrium Physics Unit, OIST Graduate University, 1919-1 Tancha, Onna, Okinawa, 904-0495, Japan}
\author{J. Samuel}
\affiliation{International Centre for Theoretical Sciences, Survey No. 151, Shivakote, Hesaraghatta Hobli, Bengaluru 560089, India}
\affiliation{Raman Research Institute, C. V. Raman Avenue, 5th Cross Road, Sadashivanagar, Bengaluru, 560080, India}
\email[]{sam@icts.res.in,sam@rri.res.in}
\author{C.-C Liu}
\affiliation{Nonlinear and Non-equilibrium Physics Unit, OIST Graduate University, 1919-1 Tancha, Onna, Okinawa, 904-0495, Japan}

\author{L. Angheluta }
\email[]{luiza.angheluta@fys.uio.no}
\affiliation{The Njord Center, Department of Physics, University of Oslo, Blindern, 0316 Oslo, Norway}
\author{A. Concha}
\email[]{andres.physics.research@gmail.com}
\affiliation{School of Engineering and Sciences,  Universidad Adolfo Ibáñez, Diagonal las Torres 2640, Peñalolen, Santiago, Chile.}
\affiliation{Theoretical Sciences Visiting Program, OIST Graduate University, 1919-1 Tancha, Onna, 904-0495, Japan} 
\altaffiliation{Condensed Matter i-Lab, Universidad Adolfo Ibáñez, Diagonal las Torres 2640, Building D, Peñalolen, Santiago, Chile.}
\altaffiliation{CIIBEC, Research Center, Santiago, Chile.}   
\author{M. M. Bandi}
\email[]{bandi@oist.jp}
\affiliation{Nonlinear and Non-equilibrium Physics Unit, OIST Graduate University, 1919-1 Tancha, Onna, Okinawa, 904-0495, Japan}

\date{\today}

\begin{abstract}
We show that surface waves in a draining-bathtub vortex provide a hydrodynamic realization of both Aharonov–Bohm phase shifts and Lense–Thirring frame dragging within a single system. A static time transformation maps the flat (2+1)-dimensional wave equation onto the convected shallow-water equation, yielding an effective vector potential set by the background flow. In this geometry, the circulation defines a global phase holonomy that controls wave structure. Traveling waves exhibit wavefront dislocations characteristic of Aharonov–Bohm scattering, while standing-wave superpositions produce nodal patterns that rotate at an angular velocity fixed by the circulation, providing a direct analogue of frame dragging. For noninteger circulation, the problem is naturally defined on the universal cover, ensuring single-valued partial-wave solutions. Experiments on a controlled vortex confirm these predictions and establish a laboratory platform in which topological phase and inertial effects, central to gauge and gravitational physics, emerge from a measurable velocity field.
\end{abstract}
\maketitle

Some of the most exotic physical effects arise not from local forces, but rather from how a system explores its global topology. A paradigmatic example is the Aharonov-Bohm (AB) effect, in which a charged particle acquires a measurable phase shift even while traveling only through a region where the magnetic field is zero, provided there is magnetic flux piercing the plane \cite{Aharonov1959}.  The phase shifts reflects a nontrivial holonomy associated with encircling the magnetic flux. A second canonical example is the Lense-Thirring (LT) effect, where a rotating mass in general relativity drags the surrounding spacetime, inducing precession of nearby trajectories, gyroscopes, or other reference frames to precess \cite{LenseThirring1918}. 
While these phenomena arise in very distinct physical systems and length-scales, both are governed by circulation: a global quantity that imprints observable phase or rotational shifts through the topology of the underlying field configuration.

Analogue systems provide controlled platforms in which topological phase phenomena can be isolated and probed experimentally. In particular, water-wave experiments have enabled studies of wave propagation, effective horizons, and phase effects arising from nontrivial background flows \cite{foglizzo2012shallow,smirnova2024water,wang2025topological,closa2010capillary}. Building on our recent study of a draining-bathtub vortex system as a hydrodynamic analogue of the AB effect for standing waves \cite{Singh2026}, we demonstrate here that the same experimental system simultaneously realizes analogues of both AB phase shifts and LT frame dragging. Surface waves on a weak, slowly varying flow satisfy a convected wave equation that can be mapped to a flat-space wave equation with an effective vector potential determined by the background circulation. In this formulation, circulation acts as a topological invariant governing wave evolution. Traveling waves exhibit wavefront dislocations characteristic of AB scattering, while standing-wave superpositions generate nodal patterns that rotate rigidly with an angular velocity set by the circulation, providing a direct hydrodynamic analogue of frame dragging.

The hydrodynamic observable is the surface deformation $\eta(t,\mathbf{r})$ of shallow water in a weak background flow $\UU(\mathbf{r})$. In the regime $|\UU|\ll c$ and with flow gradients varying on scales large compared with the ripple wavelength, $\eta$ obeys the convected wave equation
\begin{equation}
\frac{1}{c^{2}}\left(\partial_t+\UU\!\cdot\!\nabla\right)^{2}\eta=\nabla^{2}\eta
\label{eq:convected}
\end{equation}
where $c=\sqrt{gH}$ is the shallow-water wave speed, $g$ the acceleration due to gravity and $H$ the depth of water column. Outside the vortex core, the flow is approximately irrotational with the familiar azimuthal form $\UU\propto \hat{\boldsymbol\varphi}/r$ in the asymptotic region, precisely the structure underlying AB phase shift and the description of LT frame dragging.

%---------- fig1------
\begin{figure*}[t!]
    \centering
\includegraphics[width=1.0\linewidth]{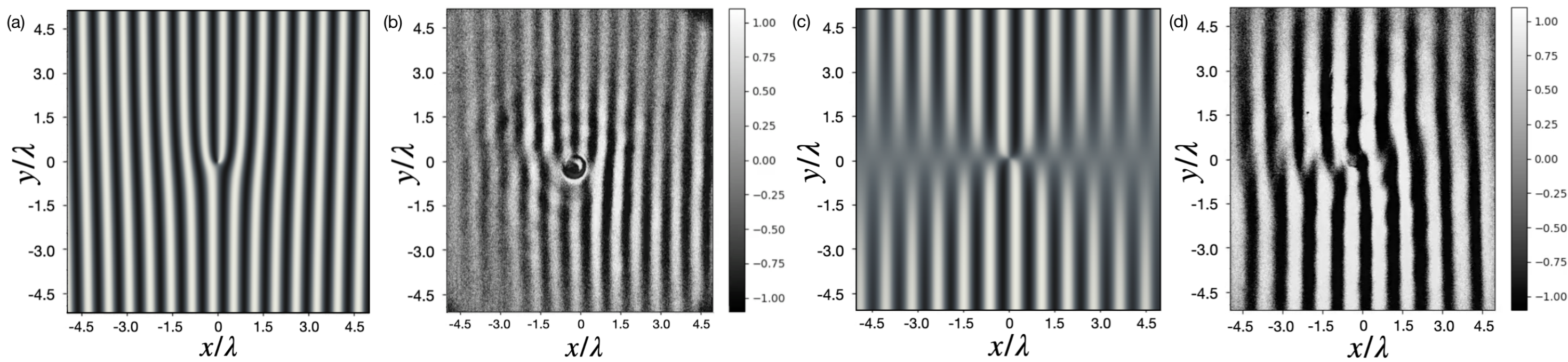}
    \caption{Traveling waves with wavefront dislocations showing map of $\eta = \textrm{Re}(\Psi)$ at $t=0$ for $\alpha=1$: (a) analytics and (b) experiment (also see S1.mov). Standing waves with nodal lines showing map of time-independent part of the $\eta= \textrm{Re}(\Psi)$ for $\alpha=1$: (c) analytics and (d) experiment (also see S2.mov in Supplementary Material).}
    \label{fig:dislocations}
\end{figure*}
%-------------------

We begin  with a complex scalar field $\Psi$ in flat $(2+1)$-dimensional spacetime with metric $\dd s^{2}=c^{2}\dd t^{2}-\dd x^{2}-\dd y^{2}$ satisfying
\begin{equation}
\square\Psi\equiv\left(\nabla^{2}-\frac{1}{c^{2}}\partial_{t}^{2}\right)\Psi=0
\label{eq:flatkg}
\end{equation}
By introducing a static time transformation,
\begin{equation}
t\mapsto \tau= t+\chi(x,y)
\label{eq:tshift}
\end{equation}
with $\chi$ independent of $t$, the metric takes the form $\dd s^{2}=c^{2}\bigl(\dd t+\partial_i\chi\,\dd x^i\bigr)^{2}-\delta_{ij}\dd x^i\dd x^j$. Transforming Eq.~\eqref{eq:flatkg} to leading order in $|\nabla\chi|\ll 1/c$ (weak-flow approximation) yields $\frac{1}{c^{2}}\bigl(\partial_\tau+c^{2}\nabla\chi\cdot\nabla\bigr)^{2}\Psi=\nabla^{2}\Psi$, with the velocity field $\UU\equiv c^{2}\nabla\chi$. Taking the real part, $\eta=\mathrm{Re}\,\Psi$, gives a solution of the convected wave equation Eq.~\eqref{eq:convected}.   
Thus, a geometrically trivial coordinate transformation in the relativistic description maps onto a physically moving medium in the fluid description (see Supplemental Material for details).

When the flow velocity is \emph{globally potential} ($\mathbf U = c^2\nabla \chi$) with a single-valued $\chi(\mathbf r)$, Eq.~\eqref{eq:convected} admits plane traveling wave solutions in the $\tau$ coordinate, $\Psi_0=\ee^{-\ii(\nu \tau-\mathbf{k}\cdot\mathbf{r})}=\ee^{-\ii[\nu t+\nu\chi(\mathbf{r})-\mathbf{k}\cdot\mathbf{r}]}$ with dispersion $\nu = c|\mathbf{k}|$. The surface deformation is  
$\eta_0=\cos\!\bigl(\mathbf{k}\cdot\mathbf{r}-\nu t-\nu\chi(\mathbf{r})\bigr)$. This case is conceptually important as it shows the background flow enters solely through a position-dependent phase shift generated entirely through the transformation Eq.~\eqref{eq:tshift}. For globally potential flows, the effect of advection is purely geometric, amounting to a phase shift of traveling waves without modifying the wave pattern.

The situation changes qualitatively when the potential $\chi$  becomes multivalued upon introducing a vortex, where the flow remains locally potential, but $\chi$ is no longer defined globally. Outside the vortex core, the velocity field is asymptotically azimuthal,
\begin{equation}
\UU=\frac{\Gamma}{2\pi r}\,\hat{\boldsymbol\varphi},
\qquad r>R,
\label{eq:vortexU}
\end{equation}
so that locally $\chi = \frac{\alpha}{\nu},\varphi$, where $\alpha \equiv \frac{\Gamma \nu}{2\pi c^{2}}$ and $\Gamma = \oint \UU \cdot d\ell$ is the circulation. Unlike the globally-potential case, the angular coordinate $\varphi$ is not single-valued, thus the phase shift $-\nu\chi(\mathbf{r})$ becomes path-dependent. Encircling the origin, the field acquires a phase $2\pi\alpha$, reflecting a nontrivial holonomy set by the circulation—this is the topological origin of the AB and LT effects.
 
A traveling plane wave in the $(\tau,\mathbf r)$ frame modified by scattering from a vortex (or phase dislocation) located at the origin, yields
\begin{equation}
\Psi_{\mathrm{tr}}(r,\varphi,t)=\eta_0\ee^{\ii(kr\cos\varphi-\nu t-\alpha\varphi)},
\label{eq:traveling}
\end{equation}
where $k=\nu/c$ and wave amplitude $\eta_0$. The factor $e^{-i\alpha\varphi}$ encodes the dislocation induced azimuthal phase winding, resulting in a multivalued phase structure in the angular coordinate. For integer $\alpha$, the field is globally single-valued under $\varphi \to \varphi + 2\pi$, and therefore constitutes a well-defined wave solution in the punctured plane $\mathbb{R}^2\setminus\{0\}$. The corresponding physical surface displacement is, $\eta_{\mathrm{tr}}=\eta_0 \cos (kr\cos\varphi-\nu t-\alpha\varphi)$, which exhibits a phase singularity at $r=0$, corresponding to the wavefront dislocation in direct analogy with the AB water-wave patterns discussed by Berry \textit{et al.} \cite{Berry1980} (Fig.\ref{fig:dislocations} (a,b)).

We now consider the standing-wave configuration formed by superposing two counter-propagating waves scattered by the vortex. The left-propagating wave along $-\mathbf k$ corresponds to spatial inversion $\mathbf r \to -\mathbf r$, under which $\varphi \to \varphi \pm \pi$ such that $\Psi_{\mathrm{tr}}^{\mathrm{left}} = \eta_0 e^{i(-kr\cos\varphi-\nu t-\alpha \varphi \mp \alpha\pi)}$. Thus, the vortex introduces a relative phase shift $\mp i\alpha\pi$ even when there is no intrinsic phase offset of the incoming plane waves. The interference of the two scattering travelling waves therefore yields a standing-wave field (see S3.mov in the Supplementary Material)
$\Psi_{\mathrm{sw}} = \Psi_{\mathrm{tr}}^{\mathrm{left}} + \Psi_{\mathrm{tr}}^{\mathrm{right}}
$, expressed as 
\begin{align}
\Psi_{\mathrm{sw}}(r,\varphi,t)
&= 2\eta_0\, e^{-i\left(\nu t + \alpha \varphi \pm  \alpha\frac{\pi}{2}\right)}
\cos\!\big( kr\cos\varphi \pm  \alpha\frac{\pi}{2} \big)
\label{eq:sw}
\end{align}
when the vortex-induced phase shift is redistributed symmetrically between the counter-propagating traveling waves. Nodal lines defined by $\eta_{\mathrm{sw}} \equiv  \textrm{Re}(\Psi_{\mathrm{sw}}) = 0$, correspond to temporal–angular nodal lines
\begin{equation}
\varphi = 
\begin{cases}
 \frac{2n+1}{2\alpha}\pi-\frac{\nu}{\alpha}t, \quad \alpha \textrm{ even}\\ 
 \frac{n}{\alpha}\pi-\frac{\nu}{\alpha}t, \quad \alpha \textrm{ odd},  
\end{cases}
\label{eq:nodal_time}
\end{equation}
where $n \in \mathbb{Z}$. These nodal lines undergo rigid rotation with angular speed $\Omega=\frac{\nu}{\alpha}$ (see Figs. \ref{fig:dislocations}(c,d), \ref{fig:phase} and S2.mov in the Supplementary Material).
This rotation is a classical hydrodynamic analogue of LT frame dragging, where the vortex circulation enforces a global, topologically constrained rotation of the standing-wave pattern, analogous to the way a rotating mass drags inertial frames in GR \cite{LenseThirring1918,Iorio2004,Pfister2007,miller2019rapidly,krishnan2020lense}. Unlike traveling waves, where the flow manifests only as a local phase shift, standing-wave nodal lines offer a direct, experimentally accessible signature of the underlying topological structure.

%-------fig 2-------
\begin{figure}[t!]
    \centering
\includegraphics[width=0.4\linewidth]{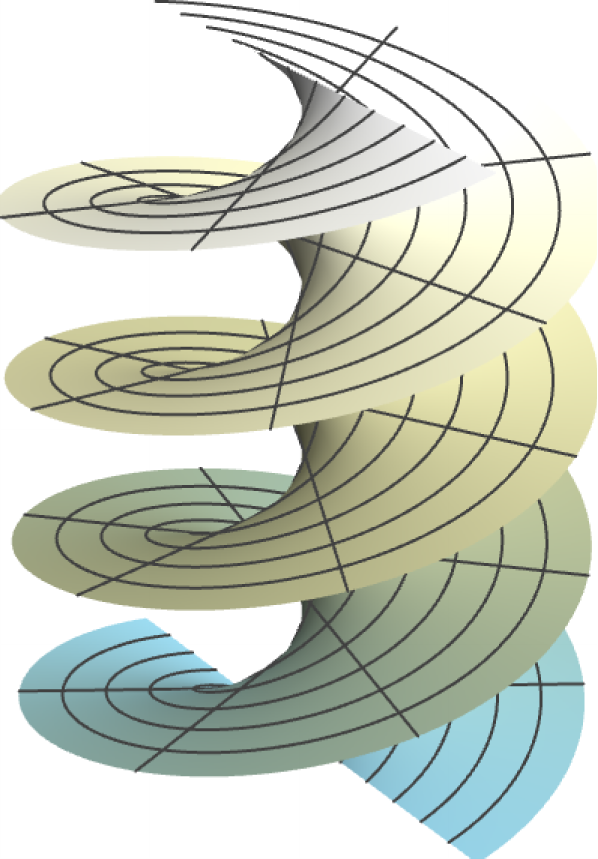}
    \caption{Topological representation of the universal cover of the punctured plane has a helicoid geometric shape.}
    \label{fig:cover}
\end{figure}
%------------
For non-integer $\alpha$, the traveling-wave solution in Eq.~\eqref{eq:traveling} acquires the monodromy
$\Psi_{\mathrm{tr}}(t,r,\varphi+2\pi)=\ee^{-2\pi\ii\alpha}\Psi_{\mathrm{tr}}(t,r,\varphi)$,
and is therefore not single-valued on the punctured plane.  This reflects the fact that the phase depends on how many times a trajectory winds around the origin. A natural way to represent this structure is to replace the angular coordinate by an “unwrapped” variable $u\in\mathbb{R}$, which keeps track of the winding number explicitly: points differing by $2\pi$ in $\varphi$ are no longer identified, but correspond to different sheets labeled by $u$ (Fig.~\ref{fig:cover}). This construction is known as the universal cover of the punctured plane and records the full winding history of the field. On the cover, we define
$\tau=t+\frac{\alpha}{\nu}u$, in terms of which the metric becomes flat $\dd s^{2}=c^{2}\dd\tau^{2}-\dd r^{2}-r^{2}\dd u^{2}$), and the wave equation separates. For
positive frequency $\nu$ one has $\widetilde{\Phi}_{\alpha}^{\nu}(\tau,r,u)=\ee^{-\ii\nu\tau}
\sum_{l\in\mathbb{Z}} \ii^{|l+\alpha|}\, \ee^{\ii(l+\alpha)u}J_{|l+\alpha|}(kr)$. Descending back to the punctured plane gives the single-valued field $\Phi_{\alpha}^{\nu}(t,r,\varphi)=\ee^{-\ii\nu t}
\sum_{l\in\mathbb{Z}} \ii^{|l+\alpha|}\, \ee^{\ii l\varphi}J_{|l+\alpha|}(kr)$. This reduces to the ordinary Jacobi--Anger expansion when $\alpha\in\mathbb{Z}$ and reproduces the standard Aharonov–Bohm partial-wave structure otherwise.

A standing-wave solution is obtained by superposing the lifted solutions propagating in opposite directions prior to projection onto the physical plane (see Supplemental Material for details). This construction highlights that the non-integer-$\alpha$ case is not a perturbative correction to the integer case, but a direct consequence of the nontrivial topology of the punctured plane; the universal cover provides the natural setting in which this structure becomes linear and globally well defined.

\begin{figure}[t!]
    \centering
\includegraphics[width=0.9\linewidth]{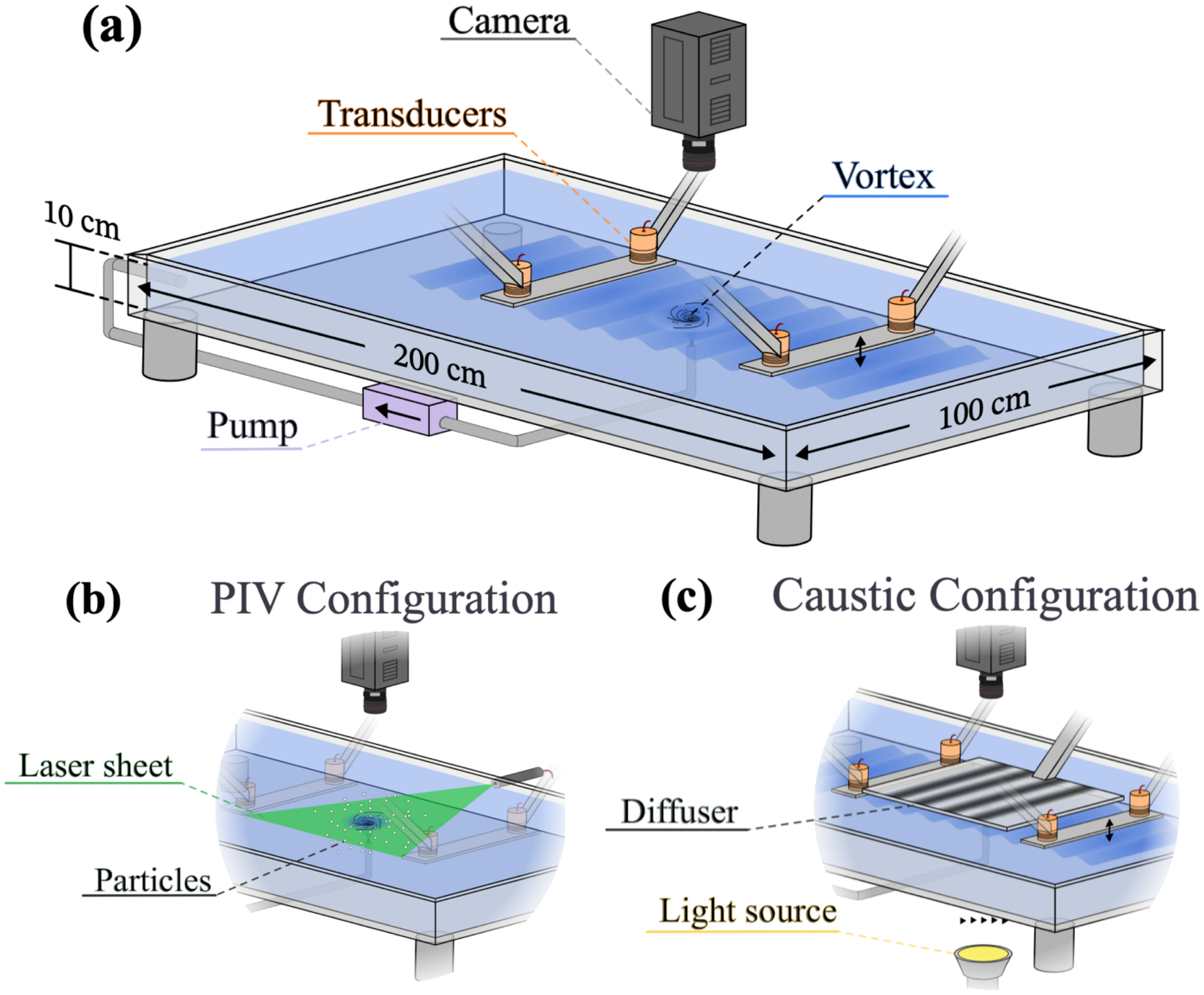}
    \caption{Schematic of the experimental setup: (a) A water tank 2 m $\times$ 1 m in dimensions is filled with water to height $H = 0.05$ m with a draining vortex generated by a pump that recirculates water back into the tank. Acoustic transducers placed on either side of the vortex generate traveling or standing surface waves. The vortex circulation is measured through Particle Imaging Velocimetry (b) whereas the wave fronts are measured with a Gaussian diffuser on to which surface caustics are projected by a backlit display. A high speed camera suspended above the tank provides imaging in both PIV and caustic configurations.}
    \label{fig:expsetup}
\end{figure}

This shared topological structure is realised in a classical hydrodynamic experiment. We consider a shallow-water hydrodynamic system operated in the weak-flow regime, where Eq.~\eqref{eq:convected} is asymptotically valid. A controlled draining-bathtub vortex is generated in a large rectangular tank of dimensions 2 m $\times$ 1 m, filled with water to a height of $H = 0.05$ m. Traveling waves are driven by acoustic transducers on one side of the vortex, while standing waves are generated using phase-locked transducers on opposite sides. The vortex circulation is tuned via the pump-driven outflow and quantitatively measured using Particle Imaging Velocimetry (PIV), allowing determination of the dimensionless circulation parameter $\alpha$. Surface deformations are visualized by a caustic-imaging technique, in which transmitted light through the free surface is focused onto a screen and recorded with a high-speed camera, producing high-contrast wavefront patterns (Fig.~\ref{fig:expsetup}). Further experimental details are given in Ref.~\cite{Singh2026}. 

%---- fig 3-------
%-------------

The coupling between the waves and the azimuthal vortex flow realizes a hydrodynamic analogue of both Aharonov–Bohm phase shifts and Lense–Thirring frame dragging. In traveling-wave experiments, this coupling produces wavefront dislocations localized near the vortex core, as originally predicted in Ref.~ \cite{Berry1980} and reproduced in the present experiments (Fig.~\ref{fig:dislocations}(a,b) and S1.mov). In contrast, standing-wave superpositions generate system-spanning nodal lines of vanishing amplitude that radiate outward from the vortex (Fig.~\ref{fig:dislocations}(c,d) and S2.mov). These structures arise from interference between counter-propagating waves in the presence of circulation and directly encode the global phase winding of the flow. Their number is set by the circulation parameter $\alpha$, taking integer values for $\alpha \in \mathbb{Z}$ and evolving continuously between adjacent integers otherwise, while their angular position evolves in time, leading to rigid rotation of the nodal pattern with angular velocity $\Omega = \nu/\alpha$ (Fig.~\ref{fig:phase} and S2.mov).

In idealized systems, the nodal lines span the whole domain, but in experiments they are truncated by finite-size and dissipative effects. This truncation arises from viscosity and capillarity, which impose an effective cutoff beyond which coherent phase structures cannot be maintained. In particular, the observed limitation can be understood in terms of a capillary-drag scale \cite{closa2010capillary}. As the nodal pattern rotates, the tangential velocity increases with radius and eventually approaches the capillary-wave speed, where capillary damping suppresses further propagation. This defines a physically meaningful outer boundary for the nodal structures, set by the competition between vortex-driven advection and capillary-wave dispersion.

%----- fig 4-----
\begin{figure}[t!]
    \centering
\includegraphics[width=0.9\linewidth]{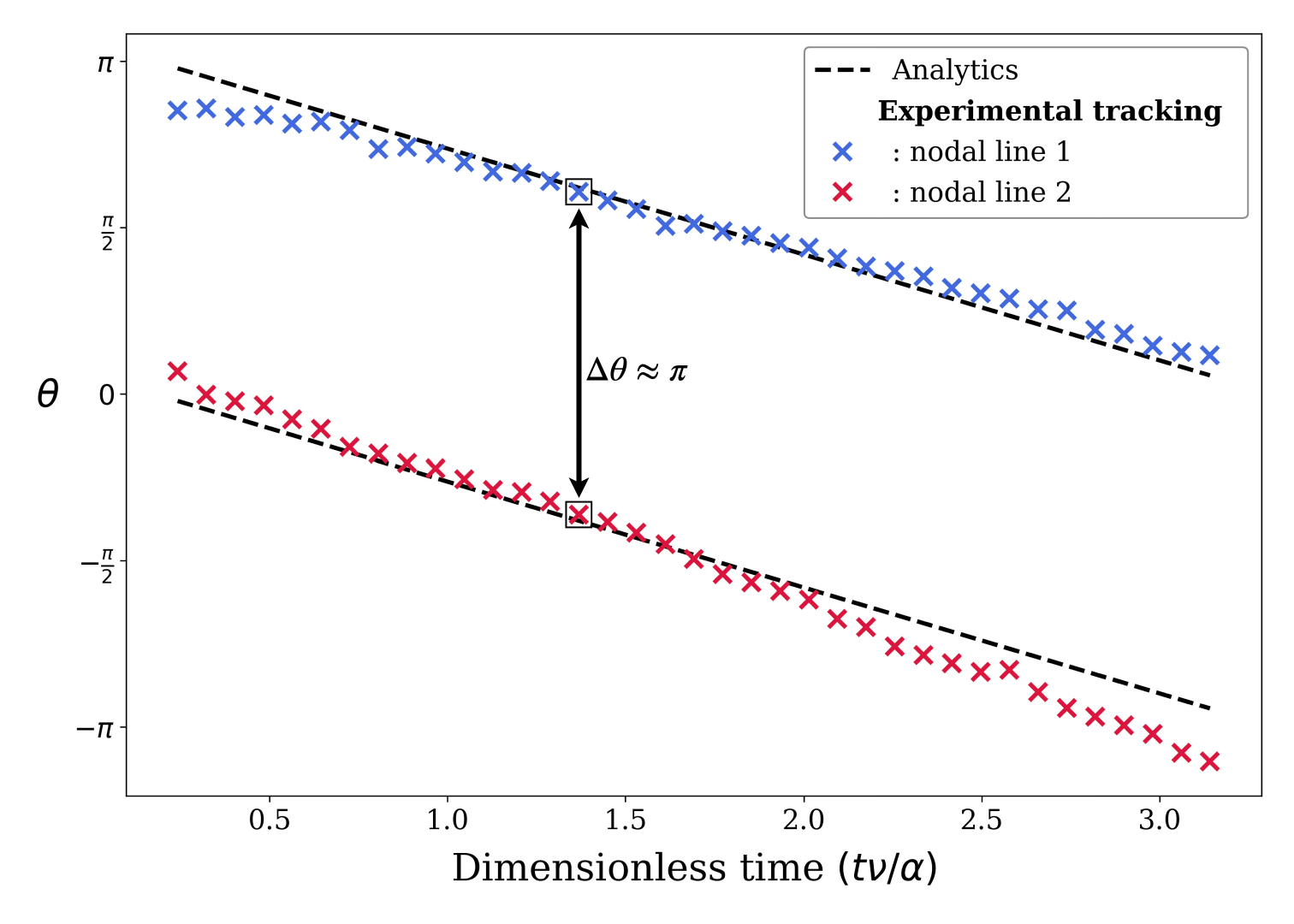}
    \caption{Phase evolution $\theta$ versus time (in units of $\nu/\alpha$) for a pair of nodal lines at $\alpha = -0.93$ obtained from analytics (dashed lines) and experiment (blue and red crosses) are separated by an azimuthal angle $\Delta \theta = \pi$ and rotate rigidly with angular velocity $\Omega = \alpha/\nu$.}
    \label{fig:phase}
\end{figure}
%---------
Taken together, these results establish a unified geometric interpretation of AB phase shifts and LT frame dragging within a single hydrodynamic analogue. In the draining-bathtub vortex experiment, both effects arise from the same underlying topological structure: a circulation-induced time transformation that generates an effective vector potential governing wave propagation. This places AB and LT phenomena on a common footing as manifestations of a single holonomy set by the vortex circulation. Unlike in quantum or gravitational settings, where the underlying vector potential or spacetime geometry is not directly accessible, the classical hydrodynamic platform allows direct measurement and control of the corresponding velocity field. The resulting wave patterns—wavefront dislocations in traveling waves and rigidly rotating nodal lines in standing waves—encode a global phase structure determined by circulation rather than local dynamics. In contrast to astrophysical realizations of frame dragging, such as those inferred from pulsar timing in white-dwarf binaries \cite{krishnan2020lense}, which are observationally constrained, the present system provides a fully tunable laboratory in which the same topological effects can be systematically probed.

This correspondence can also be viewed in a broader historical context. Early vector theories of gravity, such as those developed by Heaviside in analogy with Maxwell’s equations \cite{Heaviside1893,Maxwell1865}, anticipated a formal similarity between electromagnetic and gravitomagnetic phenomena, now encompassed by gravitoelectromagnetism. Within this framework, the AB effect is associated with electromagnetic gauge holonomy, while LT frame dragging arises from its gravitational counterpart. The present hydrodynamic system realizes this structure at the level of wave kinematics, providing a classical platform in which both effects emerge from a single vector-potential description, thus placing the observed agreement on a unified field-theoretic foundation.

More broadly, this unified perspective reveals deep connections between topology, gravitation, and quantum mechanics. LT frame dragging is most pronounced near rapidly rotating compact objects, while binary black hole mergers provide striking evidence that geometry itself can radiate through the emission of gravitational waves \cite{LenseThirring1918,krishnan2020lense,Abbott2016}. In parallel, quantum mechanics in multiply connected spaces shows that topology can enter dynamics in an equally fundamental way, through phases, interference, and inequivalent classes of motion \cite{Laidlaw1971,Osakabe1986,Leinaas1977,BleszynskiJayich2009}. These correspondences motivate extending the present hydrodynamic platform to multi-vortex configurations, where holonomy, effective gauge structure, and scattering can be explored at both geometric and dynamical levels. Our work point toward a common language in which effects usually associated with gravity, topology, and quantum mechanics emerge as different manifestations of the same underlying geometric principles.

\begin{acknowledgments}
AS, CCL, and MMB were supported through subsidy funding from the Cabinet Office of the Prime Minister of Japan to OIST Graduate University. JS acknowledges support from the Department of Atomic Energy, Government of India, under project no. RTI4019 and OIST for hospitality while part of this research was conducted. AC acknowledges partial support of FONDECYT (Chile) through grant 1250681, and by the Theoretical Sciences Visiting Program (TSVP) at OIST Graduate University.
\end{acknowledgments}

\bibstyle{apsrev4-1}
\bibliography{all}

\end{document}